\documentclass[preprint,review,12pt]{elsarticle}

\usepackage{amssymb}
\usepackage{amsmath}
\usepackage{booktabs}
\usepackage{multirow}
\usepackage{graphicx}
\usepackage{subcaption}
\usepackage{adjustbox}
\usepackage{rotating}
\usepackage{float}

\begin{document}

\begin{frontmatter}



\title{An Efficient Digital Watermarking Technique for Small Scale devices}

\author[1]{Kaushik Talathi \corref{cor1}}
\cortext[cor1]{Corresponding author}
\ead{kaushiktalathi@gmail.com}


\author[1]{Aparna Santra Biswas \corref{cor2}}
\cortext[cor2]{Principal corresponding author}

\ead{aparna.comp@coeptech.ac.in}


\affiliation[1]{organization={Department of Computer Science and Engineering },
    addressline={School of Engineering and Technology, COEP Technological University}, 
    city={Pune},
    state={Maharashtra},
    postcode={411005},
    country={India}}


\begin{abstract}
In the age of IoT and mobile platforms, ensuring that content stay authentic whilst avoiding overburdening limited hardware is a key problem. This study introduces hybrid Fast Wavelet Transform \& Additive Quantization index Modulation (FWT-AQIM) scheme, a lightweight watermarking approach that secures digital pictures on low-power, memory-constrained small scale devices to achieve a balanced trade-off among robustness, imperceptibility, and computational efficiency. The method embeds watermark in the luminance component of YCbCr color space using low-frequency FWT sub-bands, minimizing perceptual distortion, using additive QIM for simplicity. Both the extraction and embedding processes run in less than 40 ms and require minimum RAM when tested on a Raspberry Pi 5. Quality assessments on standard and high-resolution images yield PSNR $\geq$ 34 dB and SSIM $\geq$ 0.97, while robustness verification includes various geometric and signal‑processing attacks demonstrating near-zero bit-error rates and NCC $\geq$ 0.998. Using a mosaic-based watermark, redundancy added enhancing robustness without reducing throughput, which peaks at 11 MP/s. These findings show that FWT-AQIM provides an efficient, scalable solution for real-time, secure watermarking in bandwidth- and power-constrained contexts, opening the way for dependable content protection in developing IoT and multimedia applications.
\end{abstract}


\begin{highlights}
\item Propose a hybrid FWT–AQIM watermarking method optimized for small scale devices
\item Utilize YCbCr color space and apply watermark mosaic for redundancy and recovery
\item Propose a robust transform-domain features ensuring imperceptibility and resilience
\item Achieve real-time performance and strong robustness against various attacks
\end{highlights}


\end{frontmatter}



\section{Introduction}
\label{introduction}

The way we interact with media has been significantly altered by the digital revolution. Without the need for large equipment or conventional studios, people can now create, modify, and share content straight from their mobile devices, courtesy of developments in mobile technology and editing software. Users are now multidimensional participants who are simultaneously makers, consumers, and distributors as a result of this revolution. High-speed internet has further streamlined the sharing and storing of information through social media platforms and streaming services, making it easier for producers to reach large audiences \cite{Begum20, Sk18}. Digital content can be readily acquired, copied, and distributed illegally through physical transmission channels during communication, data processing, and storage. Copyright protection, content authentication, tamper detection, theft prevention, piracy, and broadcast monitoring and control are just a few of the uses for digital watermarking. By incorporating undetectable data into multimedia assets, including photographs, movies, and music, digital watermarking has become an essential method in resolving these problems by offering authenticity and protection \cite{KumarA20, Kadei22}. Furthermore, small-scale devices have become more common, especially in the context of the Internet of Things (IoT), as a result of hardware miniaturization. These little gadgets are now essential in many fields, including  home systems, industrial automation, healthcare, and military operations. Because of their extensive use, it is crucial to guarantee the safety and credibility of the digital content they process. Implementing strong security measures is made extremely difficult by these devices' constrained computational capabilities. Lightweight methods that can be easily integrated with limited hardware are crucial to closing this gap \cite{Kaw18,Reyes-Ruiz23}.

Digital watermarking involves the integration of hidden digital data into the digital content which is known as a watermark \cite{Begum20}. Digital watermarks are often divided into three main classifications: i) Multimedia based, ii) Characteristics based \& iii)Application based. Audio, pictures, video, graphics, and text are among the several media carriers in which digital watermarks are placed. Each of these media carriers has distinct signal qualities and attack surfaces that require different embedding techniques. The robustness (robust vs. fragile), visibility (visible vs. invisible), extraction method (blind vs. non-blind), embedding domain (spatial vs. frequency), and adaptivity (adaptive vs. non-adaptive) of watermarks further indicate how well they withstand processing and how noticeable they are within any carrier. These markings are used in broadcast monitoring/control, content authentication, copyright protection, and tamper detection, all of which place unique requirements on watermark fragility, detectability, and strength \cite{Yao19, Xie25, Reyes-Ruiz23}.

To put this concept into practice, each watermarking technique must go through two essential phases: i) Watermark Embedding – In this step, the watermark is embedded into the host media using a specified algorithm. ii) Watermark Detection/Extraction – In this step, the embedded watermark is checked for detectability and is then extracted from the watermarked media \cite{KumarS20}.The effectiveness of digital watermarking algorithm/s is evaluated based on three primary criteria: robustness, imperceptibility, and computational efficiency. Robustness refers to the watermark’s ability to endure various types of attacks and transformations, while the watermark should be detectable and extraction quality being acceptable. Imperceptibility is another critical criterion, ensuring that the watermark does not degrade the quality of the original content. Computational efficiency is crucial, particularly in resource-constrained environments such as IoT devices or mobile platforms. Addressing this challenge necessitates the development of lightweight, efficient security solutions tailored to the constraints of IoT devices \cite{Kapse18, Begum20}. 

The highlights of this paper can be summarized as follows:
\begin{itemize}
    \item Hybrid FWT–AQIM scheme for robust, low-distortion image watermarking.

    \item Real-time embedding/extraction on Raspberry Pi 5 with less than 40 ms latency.

    \item Supports image and QR-code watermarks up to $128 \times 128$ pixels in host images.

    \item Evaluates robustness under JPEG, noise, filtering, and geometric attacks.

    \item Achieves up to 11 MP/s throughput with less than 170 MB memory on IoT devices.
\end{itemize}

The remainder of this study is organized as follows. The related works are discussed in Section \ref{relatedWorks}. The FWT-AQIM based watermarking for small scale devices is elaborated in Section \ref{proposedMethodology}. The experimental results and analysis are presented in Section \ref{performanceEvaluation}. Finally, Section \ref{conclusion} concludes the study. 

\section{Related Works}
\label{relatedWorks}

Digital watermarking involves embedding hidden information into digital cover material in a way that allows the data to remain imperceptible yet detectable. The watermark must be resilient against standard signal processing and potential malicious attacks. It serves to uniquely identify the content owner and verify the integrity or authenticity of the carrier signal \cite{Kapse18, Gupta18}. Various watermarking systems cater to specific needs: robust watermarks are used for copyright protection; fragile or semi-fragile marks are suited for sensitive fields like medicine, forensics, intelligence, or military; and highly precise embedding is required for content authentication, where even the slightest modification is unacceptable \cite{Kapse18}.

Data hiding is achieved in both science and the arts via steganography.There are two techniques for embedding media in steganography: i) Spatial Domain and ii) Transform Domain \cite{Kadei22}. The digital watermark is directly included into the original signal’s pixel values in the spatial domain. Of all spatial domain methods, the Least Significant Bit (LSB) approach is thought to be the most straightforward. The foundation of LSB is watermarking the least significant bits of the original signal \cite{Fkirin22}. Pixel-based approaches are common in watermarking applications where real-time speed is a primary requirement for their low computational complexity and conceptual simplicity. However, they do have a number of serious shortcomings. Accurate spatial synchronization is necessary to protect users from de-synchronization attacks; neglecting the temporal axis exposes users to multiple frame collusion and video processing; and improving watermarks with only spatial analysis techniques is difficult \cite{Hussein23}. 

The spatial domain representation must first be converted into the frequency domain, and its frequency coefficients must then be adjusted, in order to insert a digital watermark in the frequency domain. Digital watermarking techniques in the transform domain include Discrete Wavelet Transform (DWT), Discrete Fourier Transform (DFT), Discrete Cosine Transform (DCT), and Singular Value Decomposition (SVD) \cite{Fkirin22}. The watermark is applied throughout the whole original data domain. The host media is translated into the frequency domain using transformation techniques. The altered domain coefficients are then used to hold the watermark data. Lastly, the watermarked media is produced using the inverse transform \cite{Hussein23}.

Recent advancements in digital watermarking have introduced various Transform domain techniques, such as the DWT and DCT, which effectively address robustness and imperceptibility. However, these methods often involve complex computations, which pose significant challenges for small-scale or resource-constrained devices. Such devices typically have limited processing power, memory, and energy resources, making it difficult to implement computationally intensive watermarking algorithms \cite{Hussein23, Panya18}.

The FWT offers a viable alternative by providing similar robustness to DWT and DCT while requiring fewer computational resources. This makes FWT a suitable choice for devices with limited capabilities \cite{Gupta18, Fkirin22}. Conversely, SVD excels in terms of robustness and imperceptibility but demands substantial computational resources, which may not be feasible for small-scale devices \cite{Kothari19}. Conventionally, QIM is more efficient in terms of computational demands but tends to compromise on robustness and imperceptibility \cite{Zareian13}.

To bridge these gaps, integrating different algorithms can yield a more balanced and efficient watermarking solution for small-scale devices. The combination of FWT and  QIM is particularly promising as it aims to reconcile computational efficiency with effective performance. This approach addresses the pressing need for practical watermarking solutions that are both robust and efficient, tailored for environments with constrained resources \cite{Begum20}. 

This study proposes a digital watermarking method that is effective and lightweight, designed for devices with limited resources. The technique strikes a balance between robustness, imperceptibility, and low computing overhead by combining Fast Wavelet Transform (FWT) with Quantization Index Modulation (QIM). This hybrid approach uses the simplicity of QIM and the efficiency of FWT to satisfy the needs of small-scale devices, in contrast to more conventional methods like DWT, DCT, or SVD, which are efficient but resource-intensive. The contribution entails refining the two techniques, creating a hybrid algorithm, and assessing its effectiveness for real-world implementation in IoT and related contexts.

\section{Watermarking based on FWT and QIM}
\label{proposedMethodology}

The proposed method, which comprises Watermark Embedding and Extraction using a
combination of FWT and an AQIM, is explained more thoroughly in this section.

\subsection{Fast Wavelet transform (FWT)}
\label{fwt}
\begin{figure}[h]
\centering
\includegraphics[width=0.75\textwidth]{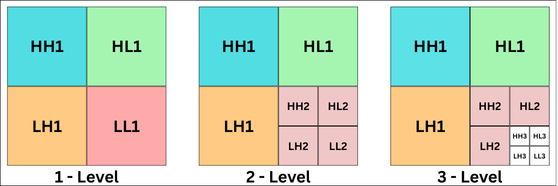}
\caption{Multiple Levels FWT Decomposition}
\label{fig:fwt-levels-demo}
\end{figure}

The FWT is a computationally efficient algorithm for performing multi-resolution signal decomposition. Low-pass and high-pass filtering are applied iteratively over rows and columns, followed by down-sampling, to transform signals from the spatial (or time) domain into the wavelet domain. Consequently, more coarse approximation and detail coefficients enable study at different scales. The original signal is reconstructed from these coefficients using synthesis filters in the inverse transform \cite{Gomes15, Arya24}. As illustrated in Fig. \ref{fig:fwt-levels-demo}, each level captures increasingly coarse image features through multi-level decomposition \cite{Vaidya22}.

\subsection{Quantization Index Modulation (QIM)}
\label{qim}

QIM is a well-known watermarking technique that maps host signal coefficients onto one of two interleaved quantization lattices, each of which represents a binary character, in order to encode information. The organized lattice arrangement guarantees resilience against common signal degradations, while the method provides imperceptibility with small perturbations of the original coefficients \cite{Zareian13}.

During the embedding process, a binary bit chooses one of the two quantization lattices that are defined by step size $\alpha$. The host coefficient is then snapped to the closest lattice point, and the bit is recovered by testing the (possibly distorted) coefficient for lattice membership later on. The selection of $\alpha$ strikes a balance between robustness (larger $\alpha$) and imperceptibility (smaller $\alpha$), which makes QIM both efficient and robust \cite{Zareian13, Kim24}.

\subsection{Watermarking processes}
\label{watermarkingProcesses}
\begin{figure}[h]
\centering
\includegraphics[width=0.75\textwidth]{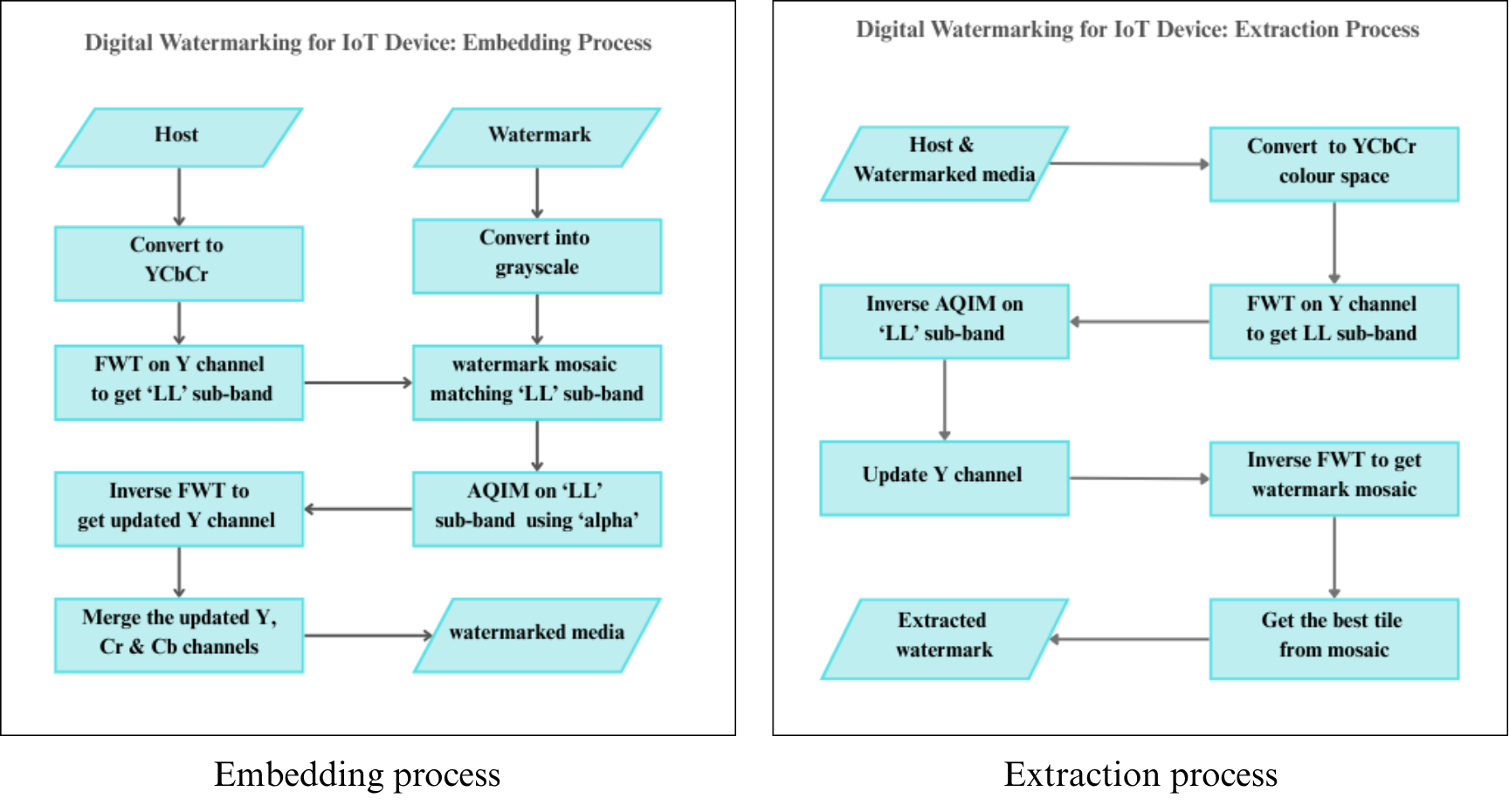}
\caption{FWT-AQIM based watermarking process}
\label{fig:watermarking-process}
\end{figure}

The proposed method supports grayscale watermarks up to $128\times128$ pixels and is compatible with any host image of size at least $512\times512\times3$. To adapt to different host–watermark combinations, users can select from multiple wavelet families and decomposition levels and adjust the quantization step $\alpha$. To minimize perceived color distortion and enable intensity-based processing, the technique transforms RGB images into YCbCr color space, a color representation that divides an image into one luminance component (Y) and two chrominance components (Cb and Cr). This is how the luminance component (Y) is calculated from RGB \cite{Yao19}. 
\begin{equation}
    Luminance (Y) = round(0.299 R + 0.587 G + 0.114 B)
\end{equation} 
The watermark mosaic is constructed in order to precisely match the low-frequency sub-band coefficient dimensions derived from the host image's Y channel's FWT. After that, the mosaic is embedded by changing the corresponding coefficients using AQIM, which maintains robustness and imperceptibility without requiring expensive norm computations by applying a controlled shift. Ultimately, the watermarked image is reconstructed using an inverse FWT. Fig. \ref{fig:watermarking-process} demonstrated the embedding process for the proposed method.

For extraction, both the original and watermarked images undergo FWT decomposition to retrieve their low-frequency coefficients. Applying the inverse AQIM rule yields a reconstructed watermark mosaic, which is then evaluated using Structural Similarity Index (SSIM) and Normalized Cross-Correlation (NCC). These metrics guide the selection of the best-matching watermark tile, ensuring accurate recovery even under common attacks. Fig. \ref{fig:watermarking-process} demonstrates the Extraction process for the proposed method.

By combining flexible wavelet settings with AQIM’s efficient additive embedding and straightforward subtraction-based extraction, the approach delivers high-quality, resilient watermarks without imposing heavy computational loads.

\section{Performance Evaluation}
\label{performanceEvaluation}

Proposed FWT-AQIM watermarking system implemented using Python on Raspberry Pi 5 having 8 GB of available memory installed with Raspberry Pi OS.  For the experiments, a common $64 \times 64$ Lena image has been used as a watermark. For the host image, various $512 \times 512$ common images like Airplane, Mandril, Peppers, Tiffany, and Cameraman are used as the host image. To widen the reach of experimentation,, an extra QR code as watermark and some non-standard uncommon large size host images are used. This system has to function under strict resource limitations, such as limited CPU, memory, and power, while maintaining data integrity and ownership protection.

\subsection{Algorithm Evaluation Metrics}
\label{evaluationMetrics}
A good evaluation system should take into account the particular performance requirements of small scale devices in addition to more conventional standards like robustness and imperceptibility. The following subsections will highlight various metrics used as part of experimentation to evaluate performance.

\subsubsection{Robustness}
\label{robustness}
Robustness is typically evaluated as the the watermark’s ability to withstand various attacks using metrics like Bit Error Rate (BER) and Normalized cross‐correlation (NCC) of the extracted watermark. A lower BER signifies that the watermark is highly resistant to attacks, whereas NCC values nearing 1 indicate almost perfect similarity.

\subsubsection{Imperceptibility}
\label{imperceptibility}
Imperceptibility evaluates the watermark's visual transparency within the host image by comparing the watermarked image and the original host image using Structural Similarity Index (SSIM) and Peak Signal-to-Noise Ratio (PSNR). Greater PSNR and SSIM values guarantee that the watermark is practically undetectable to human observers, since they show that the watermarked picture stays very close to the host.

\subsubsection{Computational Efficiency}
\label{computational-efficiency}
This parameter looks at the watermarking algorithm's efficiency in using system resources for extraction and embedding. Computational efficiency in practice includes several performance metrics:
\begin{itemize}
    \item Total time: The amount of time needed to carry out extraction or embedding,  measured under controlled circumstances to represent actual performance. How the amount of the data affects the execution time.
    \item Throughput: Usually expressed in megapixels per second (MP/s), throughput is the speed at which data (such as pictures or watermark bits) can be processed. This is essential for preserving system performance on high-demand or real-time systems.
\end{itemize}

\subsection{Base Performance analysis of Algorithm }
\label{base-performance}
Table \ref{tab:img-res} presents a unified evaluation of the algorithm’s performance across both standard and non-standard host images of variable sizes, using the Daubechies wavelet (db3) in the Fast Wavelet Transform (FWT) framework with adaptive decomposition levels (Fig. \ref{fig:fwt-levels-demo}) and quantization steps ($\alpha$) adjusted based on the watermark type. The algorithm consistently achieves high imperceptibility, with PSNR values ranging from 35 dB to over 40 dB and SSIM scores exceeding 0.99, ensuring near-lossless visual quality. Embedding the Lena watermark at $\alpha = 30$ yields PSNR around 37–38 dB and SSIM $>$ 0.99, while the QR code watermark at $\alpha = 25$ slightly reduces PSNR (34–35 dB) and SSIM (0.97–0.98), but remains visually indistinguishable. Robustness varies with watermark type—QR code watermarking exhibits near-zero BER and NCC $\approx$ 0.998–0.999 across all cases, outperforming Lena watermarking, which typically yields BER around 9–11\% and NCC $\approx$ 0.98–0.99. A notable outlier is the Tiffany image with the Lena watermark showing higher BER ($\sim$ 20\%) due to specular content, which significantly improves to 4.27\% with a QR code watermark. Computational efficiency scales with resolution, with embedding/extraction times from 30 ms to 2 seconds, throughput ranging from 2.8 to 11 MP/s, and memory usage from 157 MB to 330 MB. Despite occasional anomalies (e.g., "Test Image 2" with reduced throughput), the algorithm demonstrates excellent scalability, balancing fidelity, robustness, and resource use, making it well-suited for real-time, resource-constrained environments.

\subsection{Simulated Attack Performance}
\label{attack-performance}

The technique has experimented with a numerous popular simulated attacks, including Gaussian noise, median filter, sandpaper, compression, rotation, scaling, and cropping is covered in this section. Table \ref{tab:attack-sim} simulates algorithm performance against these attacks. 

\subsubsection{Cropping Attack}
\label{attack-cropping}
\begin{figure}[h]
  \centering
  \includegraphics[width=0.75\textwidth]{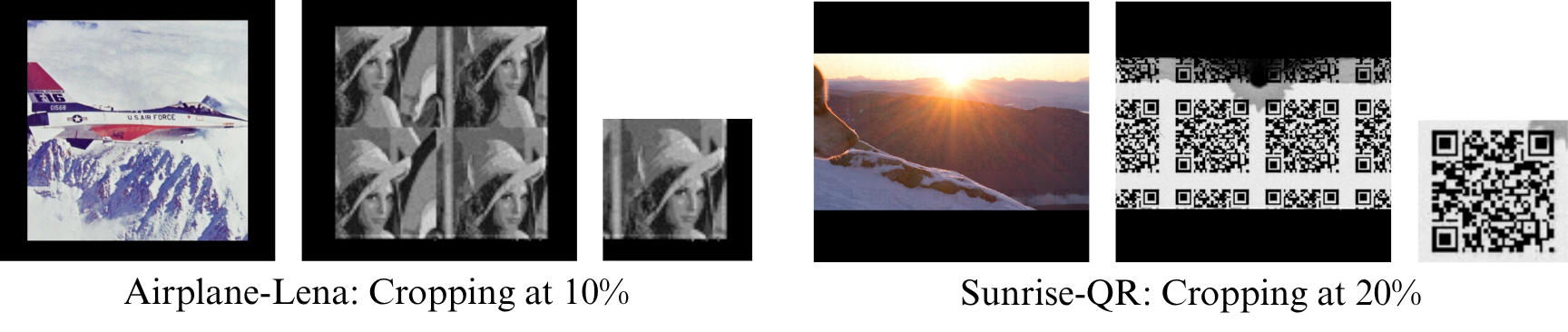}
  \caption{Illustration of the cropping attack}
  \label{fig:cropping-grid}
\end{figure}

The cropping attack simulation removes a fixed proportion \(r\) of pixels from each border and re-centers the remaining \((1-2r)\times(1-2r)\) region on a blank canvas, simulating lost or trimmed image content. For \(r\) = 10\%, overall 36\% of the image area was cropped and centered on a black canvas. Fig. \ref{fig:cropping-grid} demonstrates the cropping attack on standard and non-standard host images with different crop ratios. BER increases from around 11–15 \% to 37 \% as the crop ratio \(r\) increases from 3\% to 20\%, while NCC decreases from 0.99 to 0.28 for Lena watermarks.The QR‑code watermark shows superior robustness, with BER remaining near 0\% and NCC \(>\) 0.55 until an extreme crop ratio.

\subsubsection{Rotation Attack}
\label{attack-rotate}
\begin{figure}[h]
  \centering
  \includegraphics[width=0.75\textwidth]{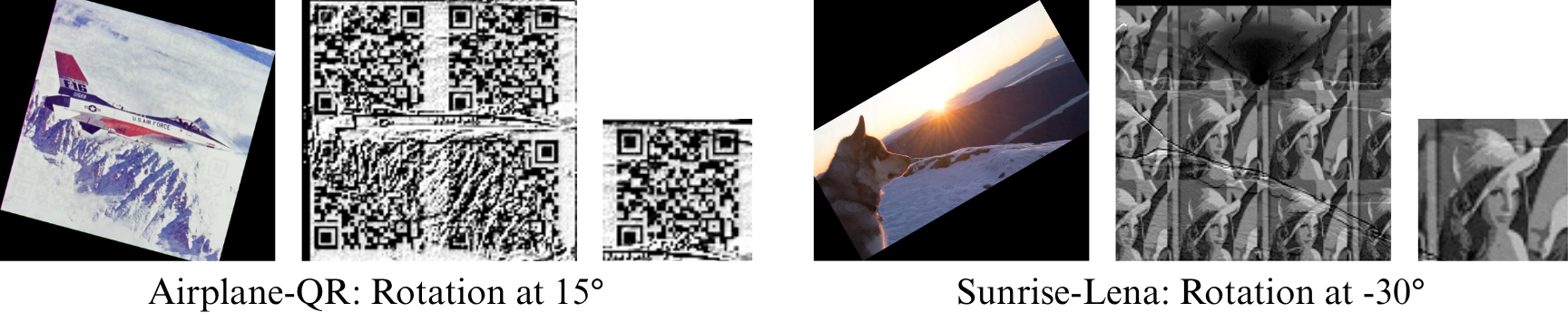}
  \caption{Illustration of the rotation attack}
  \label{fig:rotation-grid}
\end{figure}

The rotation attack turns the image by an angle $\theta$, automatically enlarging the canvas so no corners are lost. Rotation attack rotates the image by angle \(\theta\) with canvas expansion to avoid clipping of edges. Fig. \ref{fig:rotation-grid} demonstrates the rotation attack on standard and non-standard host images with different rotation angles. For Lena watermark, small rotations (±5°) causes BER rise up to 15\%, but only around 4\% for QR codes, with NCC higher than 0.9. QR codes preserves BER less than 5\% and NCC higher than 0.91 even at ±30° rotations, while Lena watermark observes BER at around 15–17\% and NCC around 0.78–0.80.

\subsubsection{Scaling Attack}
\label{attack-scaling}
\begin{figure}[h]
  \centering
  \includegraphics[width=0.75\textwidth]{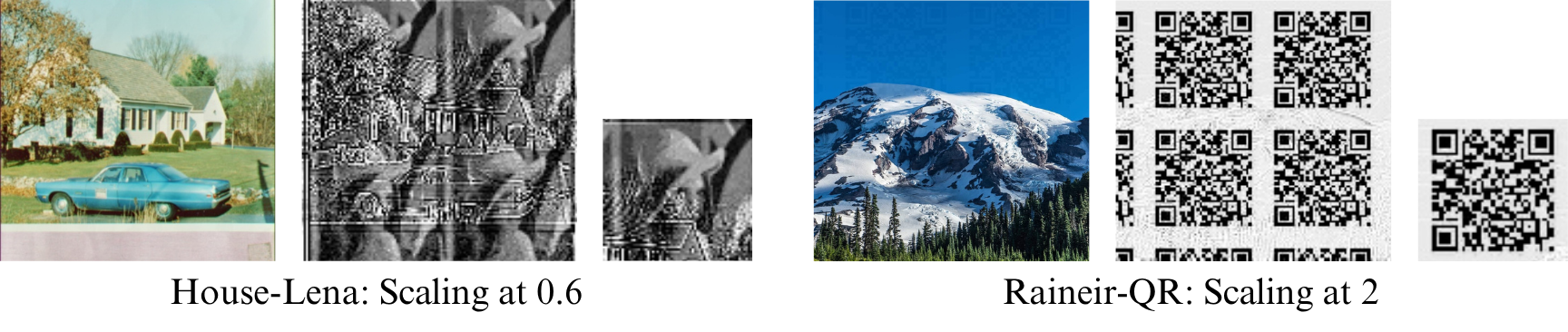}
  \caption{Illustration of the scaling attack}
  \label{fig:scaling-grid}
\end{figure}

The image is first resized by a factor \(s\) (up- or down-sampled) and then scaled back to its original dimensions using linear interpolation. This double resampling introduces smoothing, interpolation blur, and anti-aliasing artifacts similar to those seen in repeated resizing operations. Fig. \ref{fig:scaling-grid} demonstrates the scaling attack on standard and non-standard host images with different scaling factors. When down-scaling to 20\%, the NCC falls below 0.5 and the BER increases to about 35\% for Lena watermark and 23\% for QR watermark. Robustness is preserved by moderate down-scaling to 60\% and up-scaling to ×1.4, ×2 for Lena watermark at BER around 27\% and NCC higher than 0.74, for QR watermark, BER maintained less than 1\% and NCC higher than 0.98.

\subsubsection{Gaussian Noise Attack}
\label{attack-gaussian}
\begin{figure}[h]
  \centering
  \includegraphics[width=0.75\textwidth]{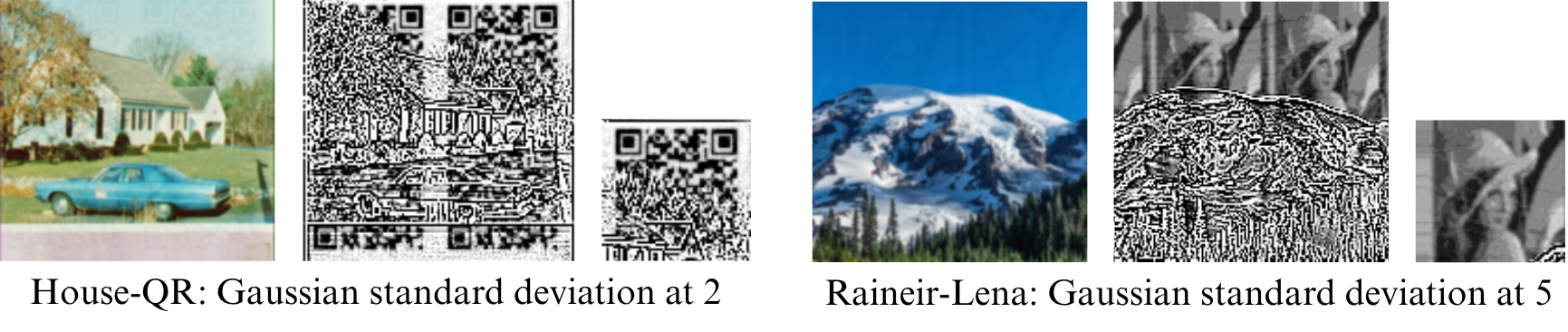}
  \caption{Illustration of the gaussian noise attack}
  \label{fig:gaussian-grid}
\end{figure}

Gaussian Noise attack injects independent, zero-mean Gaussian noise with standard deviation (\(\sigma\)) into each pixel value, modeling sensor imperfections or mild transmission disturbances. If \(\sigma\) = 5, two-thirds of pixels get shifted by at most ±5 gray levels, and about 19 out of 20 by at most ±10 levels. Fig. \ref{fig:gaussian-grid} demonstrates the Gaussian noise attack on standard and non-standard host images with different standard deviations. At \(\sigma\) = 0.5, image watermarks suffer BER at around 30\%, but QR stays at 0\% BER \& NCC around 0.99. Increasing \(\sigma\) to 10 yields BER at around 32\% for Lena watermark and up to 6\% for QR watermark, with NCC still higher than 0.89. 

\subsubsection{JPEG Compression Attack}
\label{attack-jpeg}
\begin{figure}[h]
  \centering
  \includegraphics[width=0.75\textwidth]{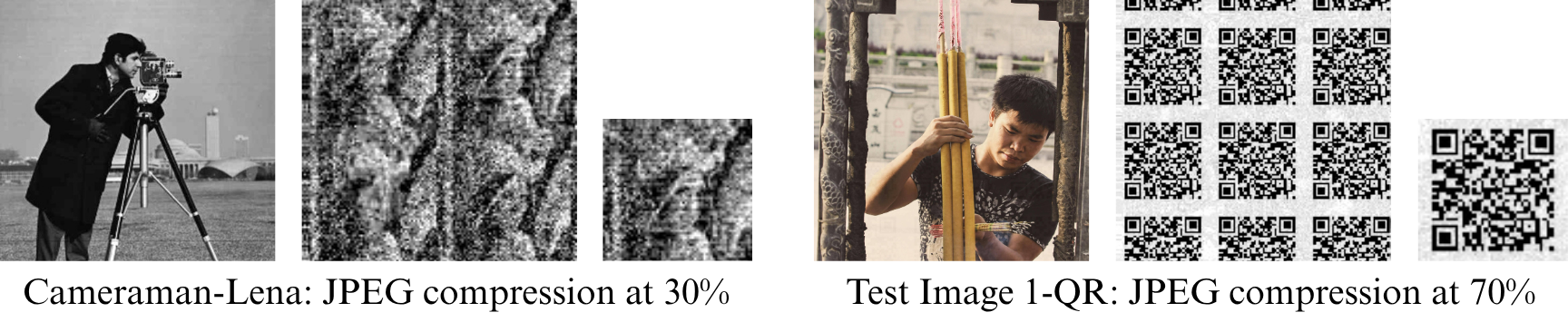}
  \caption{Illustration of the JPEG compression attack}
  \label{fig:jpeg-grid}
\end{figure}

The image is encoded and then decoded at JPEG quality \(q\). If \(q\) = 30, file size drops dramatically and visual degradation becomes very apparent. Fig. \ref{fig:jpeg-grid} demonstrates the JPEG compression attack on standard and non-standard host images with different compression values. When compressed lightly (Q = 90), the image watermark experiences about 12\% BER while the QR code remains intact at 0\% BER; both maintain NCC values exceeding 0.97. Under stronger compression (Q = 30), the QR code still recovers with only about 3–4\% BER and NCC above 0.91, whereas the Lena watermark’s BER rises to roughly 28\% and its NCC falls to about 0.65.

\subsubsection{Median Filtering Attack}
\label{attack-median}
\begin{figure}[h]
  \centering
  \includegraphics[width=0.75\textwidth]{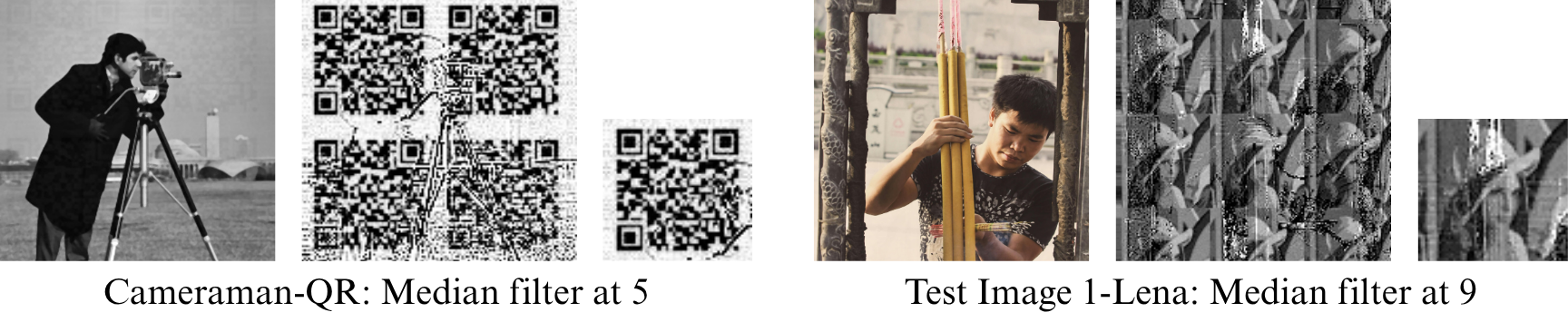}
  \caption{Illustration of the median filter attack}
  \label{fig:median-grid}
\end{figure}

Medial Filter attack applies a \(k \times k\) median filter that replaces each pixel with the median value in its neighborhood. This suppresses impulsive noise like salt-and-pepper but also softens edges and fine textures. If \(k\) is 5, the filter consists of 25 neighboring pixels for each pixel. Fig. \ref{fig:median-grid} demonstrates the median filter attack on standard and non-standard host images with different medians. A 3x3 median filter retains NCC exceeding 0.95 while producing BER values of about 16–19\% for the picture watermark and less than 1\% for the QR code. The performance of larger kernels (larger than 7×7) is drastically reduced; for watermark, BER is reaching around 30\% and for QR code, BER is reaching about 17\%, and NCC can go below 0.78.

\subsubsection{Resize Attack}
\label{attack-resize} 
\begin{figure}[h]
  \centering
  \includegraphics[width=0.75\textwidth]{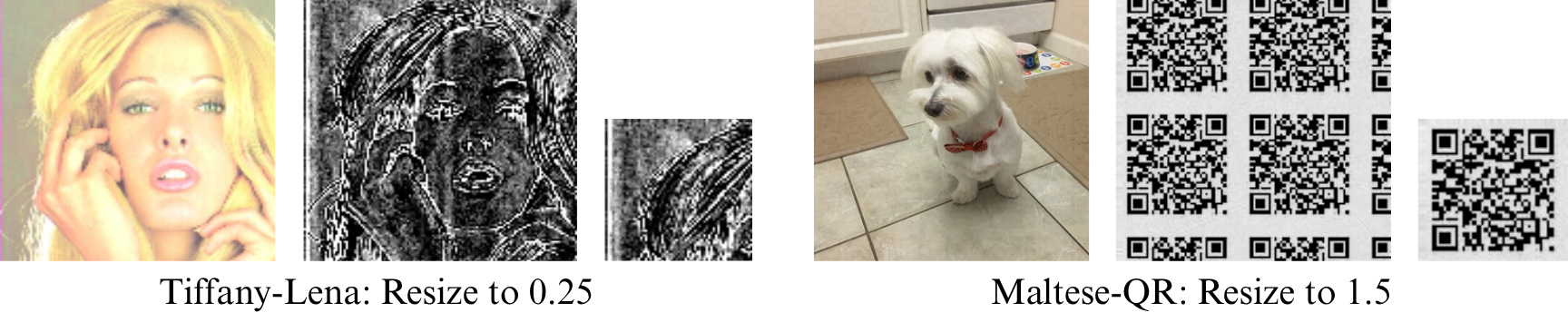}
  \caption{Illustration of the resize attack}
  \label{fig:resize-grid}
\end{figure}

Resize attack downscale image by a factor \(s\), then upscale back to its original size. If \(s\) is 5, the image shrinks to 4\% of its original area. Almost all fine details are lost while downscale. Fig. \ref{fig:resize-grid} demonstrates the resize attack on standard and non-standard host images with different resize factors. When the image is resized down to 25\%, the Lena watermark suffers approximately 27\% BER and the QR code about 19\%, with NCCs of roughly 0.54 and 0.59, respectively. In contrast, enlarging the image to 150\% limits BER to under 19\% for the Lena watermark and below 1\% for the QR code, while NCC stays above 0.90.

\subsubsection{Sandpaper Noise (Salt-and-Pepper) Attack}
\label{attack-sandpaper}
\begin{figure}[h]
  \centering
  \includegraphics[width=0.75\textwidth]{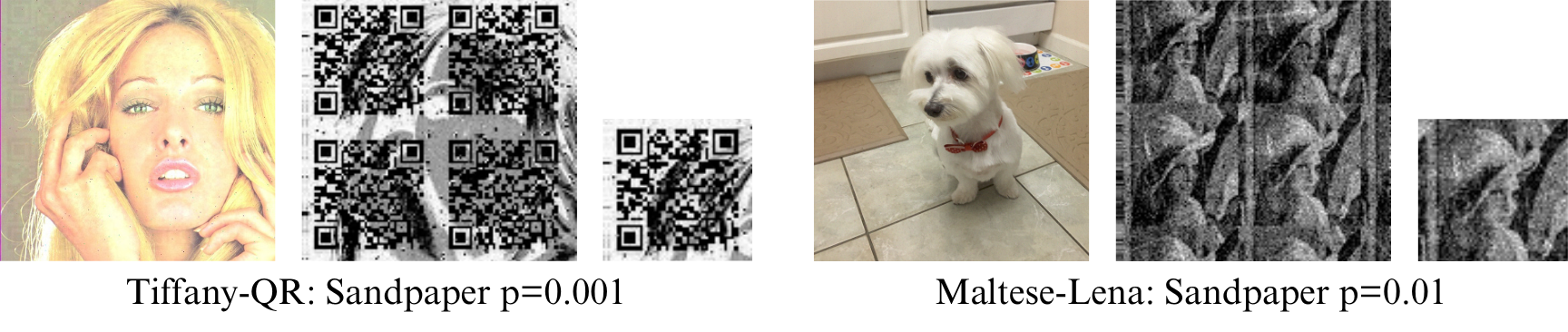}
  \caption{Illustration of the sandpaper attack}
  \label{fig:sandpaper-grid}
\end{figure}

Sandpaper attack randomly sets pixels to black or white with probability \(p\), modeling impulse noise from transmission errors or dust. For \(p)\) = 0.01, half a percent of all pixels are set to 0(zero) and the other half percent set to 1(one). Remaining 99\% of pixels are kept untouched. Fig. \ref{fig:sandpaper-grid} demonstrates the sandpaper attack on standard and non-standard host images with different probabilities. When sandpaper noise is applied at a low density (p = 0.001), the Lena watermark sees about 12–13\% BER while the QR code stays below 1\% BER, with both retaining NCCs above 0.96. Increasing the noise density to p = 0.05 causes a dramatic drop in performance. The Lena watermark’s BER surges to roughly 45\%, the QR code’s BER rises to about 34\%, and NCC falls under 0.46.

\begin{table}[tbp]
  \centering
  \caption{Algorithm simulation on standard ($512 \times 512$, level = 2, wavelet = db3) \& non-standard (variable size and parameters) Host images.}
  \label{tab:img-res}
  \begin{adjustbox}{max width=\textwidth,center}
    \begin{tabular}{@{}l *{11}{|c}@{}}
      \toprule
         & \multicolumn{2}{|c}{\textbf{Imperceptibility}} & \multicolumn{2}{|c}{\textbf{Robustness}} & \multicolumn{3}{|c}{\textbf{Embedding}} & \multicolumn{3}{|c|}{\textbf{Extraction}}  \\
        \cmidrule(lr){2-3} \cmidrule(lr){4-5} \cmidrule(lr){6-8} \cmidrule(lr){9-11}
        \centering Host Image \& parameter details & PSNR (dB) & SSIM & BER (\%) & NCC & Time (s) & Memory (MB) & TP (MP/s) & Time (s) & Memory (MB) & TP (MP/s)\\
      \midrule

      Airplane (Lena‐WM), $\alpha$ = 30
        & 37.30 & 0.9969 & 9.79 & 0.9904 & 0.031 & 157 & 8.27 & 0.032 & 163 & 8.27 \\
      \midrule

      Airplane (QR‐WM), $\alpha$=25
        & 34.53 & 0.9698 & 0.00 & 0.9984 & 0.040 & 157 & 6.44 & 0.034 & 163 & 6.44 \\
      \midrule

      House (Lena‐WM), $\alpha$ = 30
        & 37.31 & 0.9976 & 9.79 & 0.9904 & 0.039 & 157 & 8.27 & 0.032 & 158 & 6.62 \\
      \midrule

      House (QR‐WM), $\alpha$ = 25
        & 34.54 & 0.9777 & 0.00 & 0.9984 & 0.028 & 157 & 9.04 & 0.033 & 163 & 9.04 \\
      \midrule

      Peppers (Lena‐WM), $\alpha$ = 30
        & 37.35 & 0.9962 & 9.79 & 0.9962 & 0.032 & 158 & 8.13 & 0.036 & 164 & 7.34 \\
      \midrule

      Peppers (QR‐WM), $\alpha$ = 25
        & 34.56 & 0.9749 & 0.00 & 0.9983 & 0.037 & 158 & 7.01 & 0.034 & 159 & 7.71 \\
      \midrule

      Tiffany (Lena‐WM), $\alpha$ = 30
        & 38.28 & 0.9965 & 20.31 & 0.8713 & 0.032 & 158 & 8.08 & 0.034 & 164 & 7.78 \\
      \midrule

      Tiffany (QR‐WM), $\alpha$ = 25
        & 35.58 & 0.9773 & 4.27 & 0.9387 & 0.034 & 158 & 7.73 & 0.035 & 164 & 7.57 \\
      \midrule

      Cameraman (Lena‐WM), $\alpha$ = 30
        & 37.43 & 0.9914 & 9.79 & 0.9904 & 0.031 & 157 & 8.24 & 0.033 & 163 & 8.24 \\
      \midrule

      Cameraman (QR‐WM), $\alpha$ = 25
        & 34.60 & 0.9577 & 0.00 & 0.9984 & 0.032 & 157 & 7.99 & 0.033 & 158 & 7.99 \\
      \midrule
      
      Maltese (Lena‐WM, $2446 \times 2238$) ($\alpha$ = 90, Level = 4, Wavelet db3)
        & 40.39 & 0.9993 & 11.52 & 0.9801 & 0.60 & 295 & 9.18 & 0.47 & 296 & 11.62\\
      \midrule

      Maltese (QR‐WM, $2446 \times 2238$) ($\alpha$ = 90, Level = 4, Wavelet db3)
        & 38.12 & 0.9964 & 0.00 & 0.9989 & 0.60 & 294 & 9.07 & 0.49 & 295 & 11.21 \\
      \midrule

      Rainier (Lena‐WM, $1080 \times 1920$) ($\alpha$ = 60, Level = 3, Wavelet db3)
        & 37.4 & 0.9960 & 10.62 & 0.9903 & 0.23 & 206 & 8.99 & 0.202 & 207 & 10.26 \\
      \midrule

      Rainier (QR‐WM, $1080 \times 1920$) ($\alpha$ = 35, Level = 3, Wavelet db3)
        & 34.99 & 0.9905 & 0.00 & 0.9987 & 0.231 & 206 & 8.97 & 0.207 & 254 & 10.01 \\
      \midrule

      Sunrise (Lena‐WM, $2908 \times 6000$) ($\alpha$ = 90, Level = 4, Wavelet db3)
        & 40.59 & 0.9986 & 11.52 & 0.9801 & 2.195 & 252 & 7.95 & 1.916 & 330 & 9.11 \\
      \midrule

      Sunrise (QR‐WM, $1080 \times 1920$) ($\alpha$ = 70, Level = 4, Wavelet db3)
        & 38.54 & 0.9960 & 0.00 & 0.9989 & 2.175 & 252 & 8.02 & 1.900 & 330 & 9.18 \\
      \midrule

      Test Image 1 (Lena‐WM, $4390 \times 2926$) ($\alpha$ = 90, Level = 4, Wavelet db3)
        & 40.32 & 0.9988 & 11.52 & 0.9801 & 1.457 & 225 & 8.82 & 1.341 & 251 & 9.57 \\
      \midrule

      Test Image 1 (QR‐WM, $4390 \times 2926$) ($\alpha$ = 70, Level = 4, Wavelet db3)
        & 38.18 & 0.9955 & 4.27 & 0.9989 & 1.434 & 225 & 8.96 & 1.384 & 250 & 9.28 \\

      \bottomrule
    \end{tabular}
  \end{adjustbox}
\end{table}

\begin{table}[htbp]
  \centering
  \scriptsize
  \caption{Algorithm performance under various attacks on standard and non standard images}
  \label{tab:attack-sim}
  \resizebox{\textwidth}{!}{%
    \begin{tabular}{@{}l| *{8}{c|c|} @{}}
      \toprule
      & \multicolumn{4}{c|}{\textbf{Airplane}} 
      & \multicolumn{4}{c|}{\textbf{Cameraman}} 
      & \multicolumn{4}{c|}{\textbf{Sunrise}} 
      & \multicolumn{4}{c|}{\textbf{Test Img 1}} \\
      \cmidrule(lr){2-5}\cmidrule(lr){6-9}\cmidrule(lr){10-13}\cmidrule(lr){14-17}
      \multirow{2}{*}{\centering \textbf{Attack Details}} 
        & \multicolumn{2}{c|}{\textbf{Lena}} 
        & \multicolumn{2}{c|}{\textbf{QR}} 
        & \multicolumn{2}{c|}{\textbf{Lena}} 
        & \multicolumn{2}{c|}{\textbf{QR}}
        & \multicolumn{2}{c|}{\textbf{Lena}} 
        & \multicolumn{2}{c|}{\textbf{QR}}
        & \multicolumn{2}{c|}{\textbf{Lena}} 
        & \multicolumn{2}{c|}{\textbf{QR}} \\
        \cmidrule(lr){2-3} \cmidrule(lr){4-5} \cmidrule(lr){6-7} \cmidrule(lr){8-9} \cmidrule(lr){10-11}  \cmidrule(lr){12-13} \cmidrule(lr){14-15} \cmidrule(lr){16-17}
        & \textbf{BER} & \textbf{NCC}
        & \textbf{BER} & \textbf{NCC}
        & \textbf{BER} & \textbf{NCC}
        & \textbf{BER} & \textbf{NCC}
        & \textbf{BER} & \textbf{NCC}
        & \textbf{BER} & \textbf{NCC}
        & \textbf{BER} & \textbf{NCC}
        & \textbf{BER} & \textbf{NCC} \\
      
      \midrule
      
      Cropping (10\%)
        & 26.78 & 0.379 & 26.37 & 0.554  
        & 26.73 & 0.380 & 26.37 & 0.555
        & 11.52 & 0.980 & 0.00 & 0.999
        & 11.52 & 0.980 & 0.00 & 0.999 \\
      Cropping (20\%)
        & 36.89 & 0.279 & 42.06 & 0.403  
        & 36.79 & 0.280 & 42.04 & 0.405
        & 12.45 & 0.974 & 0.00 & 0.995
        & 11.52 & 0.980 & 0.00 & 0.999 \\

        \midrule
      
      Rotation (30°)
        & 15.50 & 0.795 & 4.00 & 0.911  
        & 15.26 & 0.784 & 1.85 & 0.956 
        & 12.59 & 0.953 & 0.31 & 0.994
        & 11.49 & 0.978 & 0.00 & 0.998 \\
      Rotation (-30°)
        & 16.53 & 0.762 & 4.13 & 0.910  
        & 15.38 & 0.781 & 1.88 & 0.956 
        & 12.77 & 0.953 & 0.29 & 0.989
        & 11.79 & 0.978 & 0.85 & 0.976 \\

        \midrule

      Scaling (s=0.2)
        & 34.03 & 0.405 & 22.95 & 0.555   
        & 30.59 & 0.490 & 15.53 & 0.673 
        & 32.45 & 0.831 & 4.44 & 0.926 
        & 23.71 & 0.795 & 1.81 & 0.958 \\
      Scaling (s=0.6)
        & 24.39 & 0.733 & 4.10 & 0.923   
        & 24.93 & 0.756 & 3.10 & 0.937 
        & 28.95 & 0.969 & 0.00 & 0.996 
        & 22.56 & 0.967 & 0.00 & 0.996 \\
      Scaling (s=2)
        & 12.28 & 0.915 & 0.31 & 0.987  
        & 12.35 & 0.900 & 0.61 & 0.984 
        & 10.45 & 0.979 & 0.00 & 0.999
        & 10.38 & 0.977 & 0.00 & 0.999 \\

        \midrule

      Gauss ($\sigma$=0.5)
        & 31.47 & 0.985 & 0.00 & 0.997  
        & 31.54 & 0.986 & 0.00 & 0.997 
        & 38.31 & 0.970 & 0.00 & 0.996
        & 37.31 & 0.974 & 0.00 & 0.997 \\
      Gauss ($\sigma$=5.0)
        & 27.22 & 0.829 & 0.59 & 0.966  
        & 26.95 & 0.838 & 0.59 & 0.966 
        & 36.23 & 0.954 & 0.00 & 0.991
        & 34.91 & 0.952 & 0.00 & 0.991 \\
      Gauss ($\sigma$=10)
        & 32.08 & 0.321 & 6.39 & 0.895  
        & 31.27 & 0.603 & 6.52 & 0.892 
        & 33.30 & 0.889 & 0.19 & 0.974
        & 32.22 & 0.897 & 0.19 & 0.976 \\

        \midrule
      
      JPEG (Q=70)
        & 16.58 & 0.905 & 0.00 & 0.987  
        & 17.97 & 0.891 & 0.00 & 0.986 
        & 15.45 & 0.946 & 0.00 & 0.995
        & 14.14 & 0.939 & 0.00 & 0.995 \\
      JPEG (Q=30)
        & 27.66 & 0.654 & 3.52 & 0.919  
        & 26.81 & 0.689 & 3.56 & 0.919 
        & 20.02 & 0.879 & 0.00 & 0.981
        & 16.09 & 0.876 & 0.04 & 0.981 \\

        \midrule

      Median (5x5)
        & 20.89 & 0.658 & 3.86 & 0.895  
        & 20.43 & 0.759 & 3.19 & 0.910 
        & 17.53 & 0.957 & 0.00 & 0.989
        & 15.99 & 0.943 & 0.02 & 0.995 \\
      Median (9x9)
        & 29.52 & 0.478 & 17.14 & 0.647  
        & 27.69 & 0.538 & 13.52 & 0.719 
        & 17.16 & 0.928 & 0.24 & 0.983
        & 16.26 & 0.828 & 1.39 & 0.966 \\

        \midrule

      Resize (S=0.25)
        & 31.23 & 0.439 & 19.38 & 0.588  
        & 27.29 & 0.536 & 11.49 & 0.724 
        & 25.98 & 0.869 & 0.61 & 0.965
        & 27.56 & 0.874 & 0.98 & 0.974 \\      
      Resize (S=0.75)
        & 21.17 & 0.841 & 2.32 & 0.962  
        & 20.19 & 0.872 & 1.90 & 0.965 
        & 26.39 & 0.972 & 0.00 & 0.997
        & 25.44 & 0.967 & 0.00 & 0.996 \\
      Resize (S=1.5)
        & 18.55 & 0.915 & 0.49 & 0.985  
        & 16.69 & 0.903 & 0.81 & 0.983 
        & 22.46 & 0.979 & 0.00 & 0.999
        & 15.79 & 0.974 & 0.00 & 0.998 \\

        \midrule
        
      Sandpaper (P=0.005)
        & 21.17 & 0.726 & 4.29 & 0.919  
        & 20.19 & 0.702 & 3.73 & 0.929 
        & 16.55 & 0.969 & 0.00 & 0.989
        & 16.16 & 0.868 & 0.56 & 0.978 \\
      Sandpaper (P=0.01)
        & 25.85 & 0.562 & 8.06 & 0.851  
        & 25.27 & 0.569 & 7.54 & 0.856 
        & 15.69 & 0.836 & 0.15 & 0.977
        & 18.97 & 0.772 & 1.95 & 0.949 \\
      Sandpaper (P=0.05)
        & 45.04 & 0.155 & 33.59 & 0.455  
        & 40.79 & 0.239 & 26.44 & 0.511 
        & 38.94 & 0.296 & 18.09 & 0.651
        & 30.96 & 0.433 & 22.48 & 0.567 \\
      \bottomrule
    \end{tabular}%
  }
\end{table}

\begin{figure}[h]
  \centering
  \includegraphics[width=\textwidth]{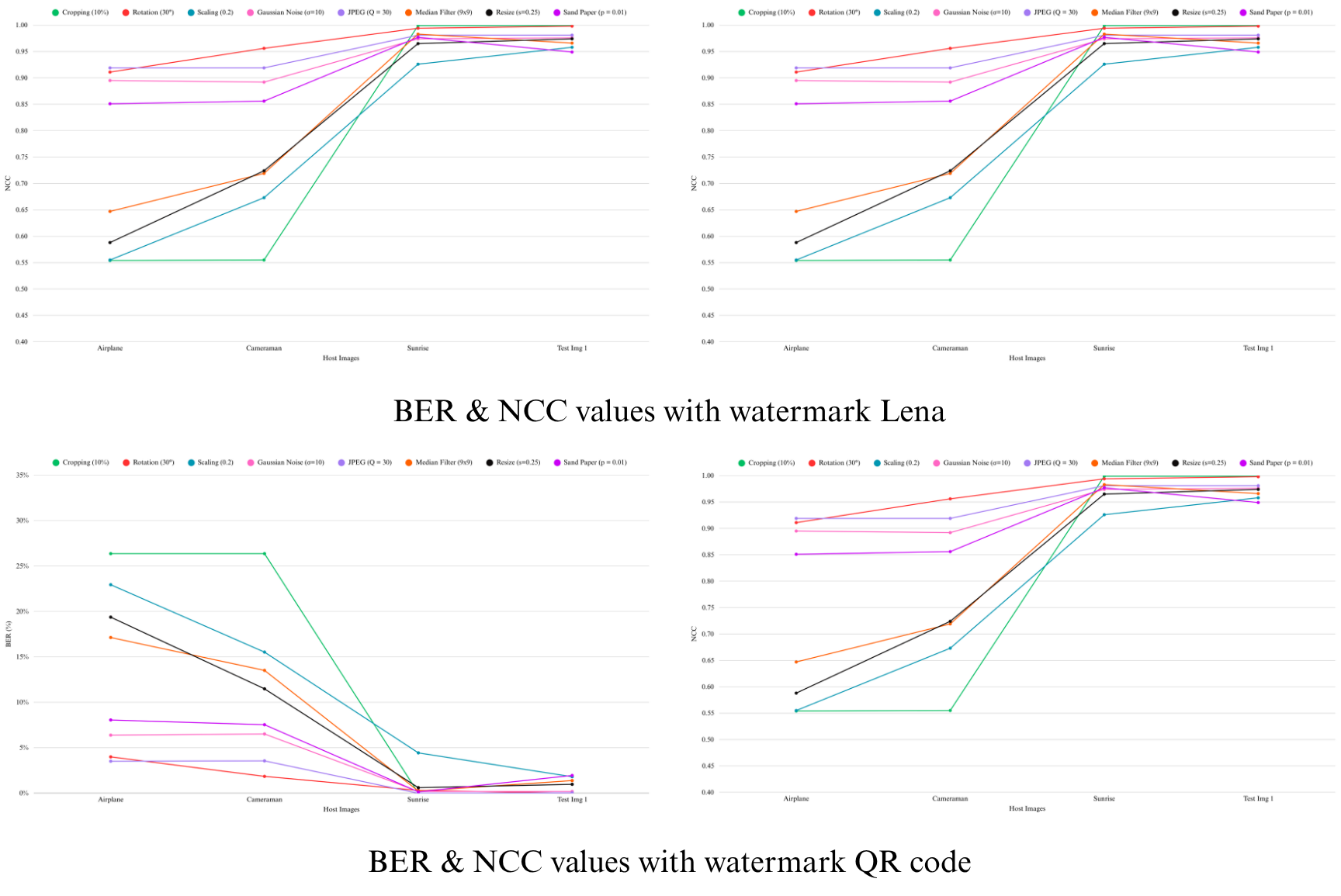}
  \caption{\centering Graphical representation of various attacks for standard and non-standard images}
  \label{fig:result-graph}
\end{figure}

\begin{table}[htbp]
    \centering
    \caption{Robustness comparison between proposed method and existing methods based on NCC values}
    \label{tab:ncc-comparison}
    \resizebox{\textwidth}{!}{%
    \begin{tabular}{@{}l  cc  cc c@{}}
        \toprule
        \textbf{Attacks} & \textbf{Proposed Method} & \textbf{Sk A, et al.\cite{Sk18}} & \textbf{Kumar Shrivastava S, et al.\cite{KumarSS18}} & \textbf{Li Z, et al. \cite{Li21}} \\
        \midrule 
        Cropping         & 0.999 & 0.9973 & 0.9869 & 0.9820\\
        Rotation         & 0.997 &  -     & 0.8562 & 0.9682\\
        Scaling          & 0.993 &  -     & 0.7212 & 1.0000\\
        Gaussian Noise   & 0.997 & 0.7114 & 0.7523 & 0.9676\\
        JPEG Compression & 0.999 &  -     & 0.7823 & 1.0000\\
        Median Filter    & 0.992 & 0.8518 & -      & 0.9908\\
        Sand paper       & 0.995 & 0.9485 & -      & 0.9350\\
        \bottomrule
    \end{tabular}
    }
\end{table}

\begin{table}[htbp]
  \centering
  \caption{Watermark embedding and extraction time (in seconds) for the proposed method and existing method for IoT devices}
  \label{tab:timing-comparison}
  \begin{adjustbox}{max width=\columnwidth,center}
  \begin{tabular}{@{}l  |c|c|c|c@{}}
    \toprule
    \multirow{2}{*}{\textbf{\shortstack{Standard\\images}}}
      & \multicolumn{2}{c|}{\textbf{Proposed Method}} 
      & \multicolumn{2}{c}{\textbf{Sk A, et al.\cite{Sk18}}} \\
    \cmidrule(lr){2-3} \cmidrule(l){4-5}
        & \textbf{Embedding} & \textbf{Extraction} & \textbf{Embedding} & \textbf{Extraction}\\
    \midrule
    Peppers    & 0.032 & 0.036 & 0.056892 & 0.029547 \\
    Baboon     & 0.033 & 0.034 & 0.020165 & 0.034652 \\
    Aeroplane  & 0.031 & 0.032 & 0.048915 & 0.025395 \\
    \bottomrule
  \end{tabular}
  \end{adjustbox}
\end{table}
 
Across all experiments—from baseline embedding tests (Table \ref{tab:img-res}) to simulated attack scenarios (Table \ref{tab:attack-sim})—the proposed FWT–AQIM watermarking scheme consistently demonstrates a well-balanced trifecta of imperceptibility, robustness, and computational efficiency across both standard and non-standard host images. Utilizing the Daubechies wavelet (db3) with adaptive decomposition levels and embedding in the luminance (Y) channel of the YCbCr color space, the method ensures high energy compaction and visual quality preservation. Embedding a $64 \times 64$ Lena watermark yields PSNR values around 37–38 dB, SSIM $>$ 0.99, BER around 9–12\%, and NCC $\approx$ 0.99, while QR code watermarking offers slightly lower PSNR (34–35 dB) and SSIM (0.97–0.98), but achieves near-zero BER and NCC up to 0.998—all under 40 ms, within 160 MB memory, and at throughput ranging from 7–11 MP/s. Under attack conditions, QR code watermarks maintain superior robustness (BER $<$ 5\%, NCC $>$ 0.90), even under severe distortions, while Lena watermarks degrade more noticeably (BER up to 42\%, NCC $\approx$ 0.75). Similar resilience is noted across high-resolution, non-standard hosts where deeper FWT levels and more mosaic tiles enhance embedding redundancy. QR codes remain highly robust (BER near zero, NCC $>$ 0.94), whereas Lena BER increases to $\sim$20\% under moderate distortion. Noise resilience is also favorable: at low sandpaper noise (p = 0.001), Lena BER is $\sim$12\% vs. $<$1\% for QR; under extreme noise (p = 0.05), both degrade, but QR remains more robust. Fig. \ref{fig:result-graph} illustrates these trends, showing how larger hosts support higher FWT levels and improved embedding granularity, especially benefitting structured watermarks like QR codes. Timing results illustrated in Table \ref{tab:timing-comparison} indicate our method completes embedding/extraction 0.03–0.04s faster than Sk A et al. \cite{Sk18}, while maintaining high NCC values ($\geq$ 0.992) across all seven attack types (Table \ref{tab:ncc-comparison}), including near-perfect recovery under scaling and JPEG compression. Altogether, FWT–AQIM offers a scalable, lightweight, and resilient watermarking solution, ideal for real-time, resource-constrained applications such as IoT systems.
 
\section{Conclusion and Future scope}
\label{conclusion}

In this study, the FWT–AQIM watermarking scheme maintains an ideal balance of speed, strength, and invisibility—exactly what is needed for devices with limited memory and power. The watermark is hidden in the low-frequency layer of the luminance channel (Y) in YCbCr space, allowing embedding and extraction of data in less than 40 ms on a Raspberry Pi 5 while minimizing visual distortion (PSNR $\geq$ 34 dB, SSIM $\geq$ 0.97). The mosaic-based spread that exploit spatial redundancy, QR-code watermarks appear nearly faultless (NCC $\geq$ 0.998) even when images undergo significant cropping, rotation, scaling, compression, noise, or filtering. With a peak throughput of 11 MP/s across a range of image sizes, it surpasses similar techniques without compromising fidelity. These findings demonstrate that FWT-AQIM based method is not only workable in low-power, real-time settings but also resilient enough to safe-guard digital content.

In the future, the system might be expanded to include adaptive quantization parameter adjusting based on local texture complexity may enhance the transparency and resilience of watermarks even further. It may be possible to further minimize localized distortions and enhance watermark clarity under specific degradations by using mosaic-based watermark reconstruction, in which all eligible tiles over a predetermined threshold are superimposed to create a single representative tile.

\section*{Declaration of Competing Interest}
The authors declare that they have no known competing financial interests or personal relationships that could have appeared to influence the work reported in this paper.%

\section*{CRediT authorship contribution statement}

\textbf{Kaushik Talathi}: Conceptualization, Methodology, Formal analysis, Data curation,  Software, Validation, Visualization, Writing – original draft. \textbf{Aparna Santra Biswas}:
Conceptualization, Methodology, Formal analysis, Visualization, Investigation, Writing – review \& editing, Supervision.

\section*{Data availability}
The data will be available as on request.

\end{document}